\documentclass{ceab}   

\usepackage{epsfig}     
\usepackage{graphicx}   

\usepackage{ceabbib}     
\usepackage[T1]{fontenc}

\begin{document}

\newcommand{\mps}{m\,s$^{-1}$}
\newcommand{\bvec}[1]{ \mbox{\boldmath$#1$} }
\def\cK{{\cal K}}

\title{An Average Supergranule: Much Larger Vertical Flows Than Expected}

\author{M. \v SVANDA$^{1,2}$
\vspace{2mm}\\
\it $^1$Astronomical Institute, Academy of Sciences of the Czech Republic (v. v. i.), \\
\it Fri\v{c}ova 298, CZ-25165 Ond\v{r}ejov, Czech Republic\\
\it $^2$Charles University in Prague, Faculty of Mathematics and Physics, \\
\it Astronomical Institute, V Hole\v{s}ovi\v{c}k\'ach 2, CZ-18000 Prague 8, Czech Republic
}

\maketitle

\begin{abstract}
Supergranules are believed to be an evidence for large-scale
subsurface convection. The vertical component of the supergranular flow
field is very hard to measure, but it is considered only a few \mps{} in and
below the photosphere. Here I present the results of the analysis using three-dimensional inversion for time-distance
helioseismology that indicate existence of the large-magnitude vertical upflow in the near sub-surface layers.
Possible issues and consequences of this inference are also discussed.
\end{abstract}


\section{Supergranules: Convection cells?}
The photosphere of the Sun is fully covered with supergranules, a convection-like cellular pattern. Some properties of supergranules are still being debated in literature, despite the discovery of this phenomenon more than half century ago \citep{1954MNRAS.114...17H}. Surface properties of supergranules were extensively studied by many authors, e.g., \cite{1989SoPh..120....1W}, \cite{1998ASPC..140..161S}, \cite{2004ApJ...616.1242D}, \cite{2008SoPh..251..417H} and many others. Recently, the findings about supergranulation were reviewed by \cite{lrsp-2010-2}. 

The main lesson from dozens of papers published over almost six decades is that only the surface properties of supergranules are pretty well known. The characteristic cell size is around 30~Mm (the corresponding velocity spectrum scale-length is around 38~Mm however),  lifetime of supergranules is a day or more. Supergranules cannot be observed in white light as the temperature within the cell varies by less than 4~K. 
The plasma flows within the supergranules are predominantly horizontal with root-mean-square velocity  
$\sim$300~\mps, thus supergranular pattern is easily observable in Dopplergrams. The vertical (radial, with respect to the spherical Sun) component of the supergranular flow is hard 
to measure, its amplitude is usually within measurement error 
bars. Statistically it has been determined that the root-mean-square vertical velocity ranges from 4~\mps{} \citep{2010ApJ...725L..47D}
to 29~\mps{} \citep{2002SoPh..205...25H}. 

The depth structure and origin of the supergranulation is practically unknown. Various local helioseismology methods \citep[for a review see][]{2005LRSP....2....6G} attempted to study the sub-surface structure of supergranular flow. Some studies analysed the flow maps at various depths inferred by helioseismic inversions. They investigated correlation of horizontal divergence of the flow at various depths with the surface. This approach was used e.g. by \cite{1998ESASP.418..581D}, \cite{2003ESASP.517..417Z}, or \cite{2009NewA...14..429S} to name a few. Although the numerical values differ for individual studies, a general conclusion is that supergranules are convection-like cells extending to depths of 8--25~Mm with a deep ``return flow''. \cite{2011NewA...16....1Z} attempted to segment supergranules as compact objects in the 4-D space (three for spatial coordinates and one for time) from the sequence of helioseismic flow maps. The authors tried to use the large sample of 3-D supergranules with their histories to capture the evolution of an average supergranule in the time--depth domain. These results again support supergranules to be two-layer structures loosing the coherence at the depth of some 25~Mm or so. Note that all studies cited in this paragraph are based on RLS helioseismic inversions under ray approximation (geometrical optics) applied to phase-speed filtered data and obtained using practically the same code. 

Independent studies, employing  ridge-filtered data with Born approximation wave kernels and OLA inverse methods, e.g. \cite{2007ApJ...668.1189W} and \cite{2008SoPh..251..381J}, did not detect the flow reversal. Moreover, they pointed out that any inversion for the flow snapshot deeper than 4--6~Mm is dominated by random noise and thus does not reveal any information about the deep supergranular flow. The statement about the noise was also confirmed by \cite{Svanda2011} using data from numerical simulation of solar convection. 

Non-helioseismic methods were also used to draw conclusions about the depth structure of supergranules. E.g., \cite{2012ApJ...749L..13H} recently showed that supergranules may extend to depths equal to their widths indicating that analysis of supergranules with differing sizes may lead to a different depth structure.

Last but not least: a non-standard helioseismic method was used to investigate the vertical component of the supergranular flow just under the surface. 
A systematic offset in the travel-time shifts measured by \cite{2012SoPh..tmp..136D} for large separations between the measurement points using a special spatio-temporal filtering of the data served as evidence indicating a large-amplitude flow under the surface of supergranules. The non-standard travel-time measurement procedure was selected to avoid the cross-talk between the horizontal and vertical 
flow in supergranules, which was known to be hard to avoid in inverse modelling. Using a 
simple Gaussian model constrained by surface measurements, they predicted a peak in the 
vertical flow of 240~\mps{} at a depth of $2.3$~Mm.  

\section{Inverse modelling}
\label{sect:inversion}
I aimed to verify the results of \cite{2012SoPh..tmp..136D} using a standard time--distance inversion pipeline (described and tested in \citealt{Svanda2011}). The pipeline implements Subtractive Optimally 
Localised Averaging \citep[SOLA;][]{1992AA...262L..33P} three-dimensional inversion for time--distance \citep{1993Natur.362..430D} helioseismology applied to a set of travel-time maps. 

The travel times were measured from twelve-hour Dopplergram datacubes acquired by Helioseismic and Magnetic Imager (HMI; \citealt{2012SoPh..275..207S}, 
\citealt{2012SoPh..275..229S}) in the period of 10 June to 10 July 2011, thus resulting in 64 consecutive sets of travel-time maps. In each datacube, only the disc-centre region (512$\times$512 pixels with pixel size of 1.46~Mm) was tracked using the code {\sc drms\_tracking} (Schunker \& Burston, unpublished) deployed at German Science Center for SDO. The Dopplergram datacubes were spatio-temporally filtered with ridge filters, separating $f$, $p_1$, and $p_2$ modes. The linearised travel times were measured following \cite{2004ApJ...614..472G}, using centre-to-annulus and centre-to-quadrant geometries \citep{1997ASSL..225..241K} with radii of the 
annuli 5 to 20 pixels. 

The SOLA inversion was performed in a Cartesian approximation ($x$ and $y$ for the horizontal coordinates and $z$ for the vertical) using sensitivity kernels computed in Born approximation \citep{2007AN....328..228B} for geometries consistent with travel-time measurements (ridge-filtered modes $f$ to $p_2$ and centre-to-annulus and centre-to-quadrant averaging with same set of annuli radii). Only the near-surface layers were targeted. Note that since only the patch around the disc centre is used for the inverse modelling (to prevent effects of foreshortening, which is not accounted for in the sensitivity kernels), approximation by Cartesian coordinates is justified. The vertical velocity then almost coincides with the line-of-sight velocity (the angle between the two is less than 24 degrees), except for the oposite sign. 

The SOLA algorithm minimises a cost function consisting of many regularisation terms. Among those, important 
terms evaluate the localisation in the Sun (an averaging kernel), the level of random noise in the 
results, and cross-talk contributions. By choosing values of the trade-off parameters, one can 
regularise strongly about some of these terms, while relaxing demands on the others. 
 
Due to the random-noise realisations, a normal tomography approach cannot be used to measure the vertical component of the supergranular flow \citep{Svanda2011}. In such case, the signal-to-noise ratio is usually in the order of unity, i.e., the resulting maps are very noisy. Even in such case, the averaging kernel has many side-lobes which introduce unwanted artefacts. The horizontal side-lobes are especially dangerous. Despite being ``almost negligible'' in magnitude, they accumulate a large volume and thus represent a significant component of the total volume integral over the whole averaging kernel. 
To avoid such contamination, one has to pick the solution with good localisation, no side-lobes, and therefore higher noise level. Such estimate of the vertical flow would be completely buried in random noise. 

Luckily, independent flow maps may be considered as independent realisations of the flow, including the realisation of random noise. When one averages over $N$ these independent realisations, one increases the signal-to-noise ratio $\sqrt{N}$-times. That is the approach governing this study. It is fair to note that same approach was used also by \cite{2010ApJ...725L..47D} and \cite{2012SoPh..tmp..136D}.

\begin{figure}[!th]
\centering
\begin{tabular}{ll}
\raisebox{5cm}{$v_x$:} & \includegraphics[width=0.9\textwidth]{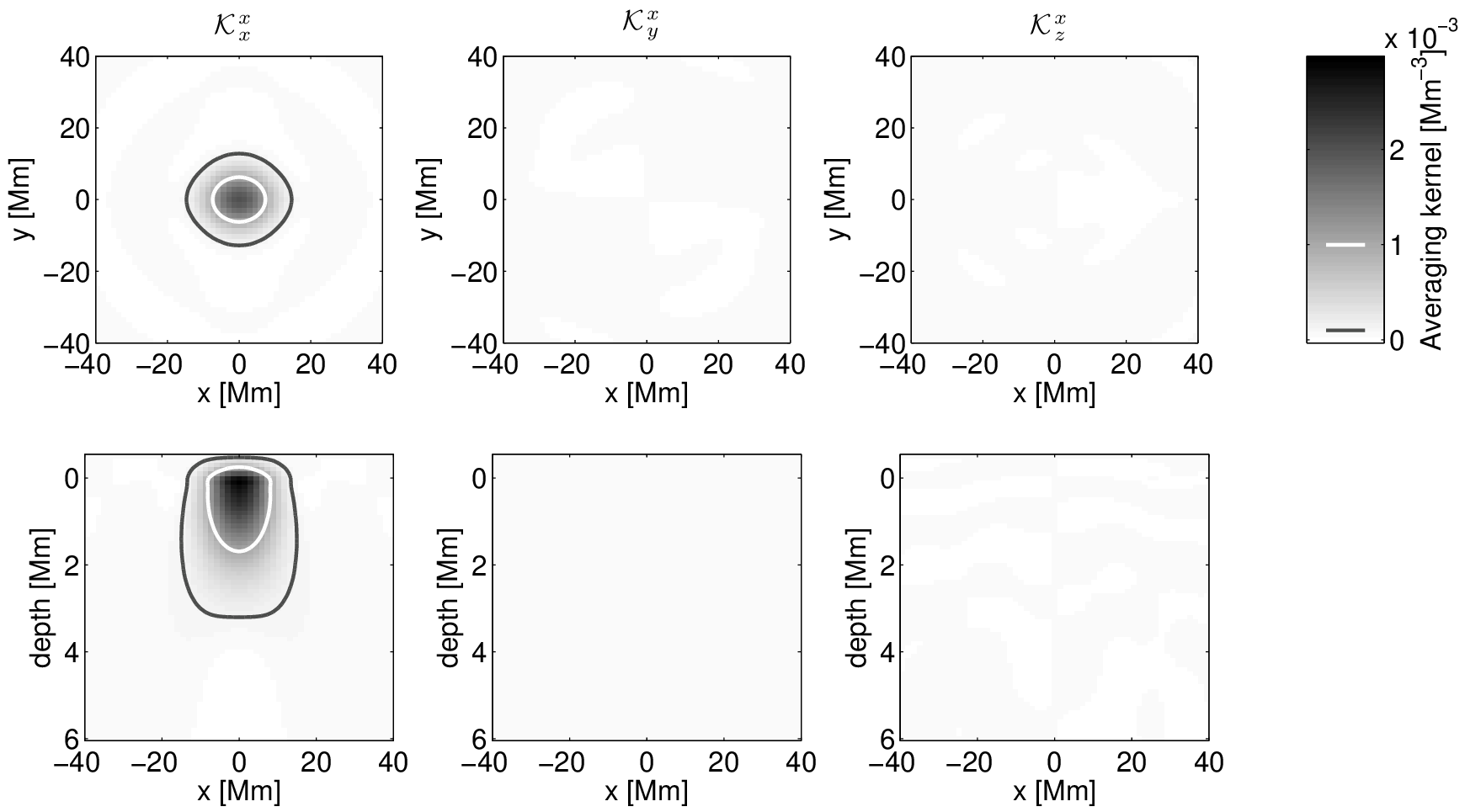}\\
\hline
\raisebox{5cm}{$v_z$:} & \includegraphics[width=0.9\textwidth]{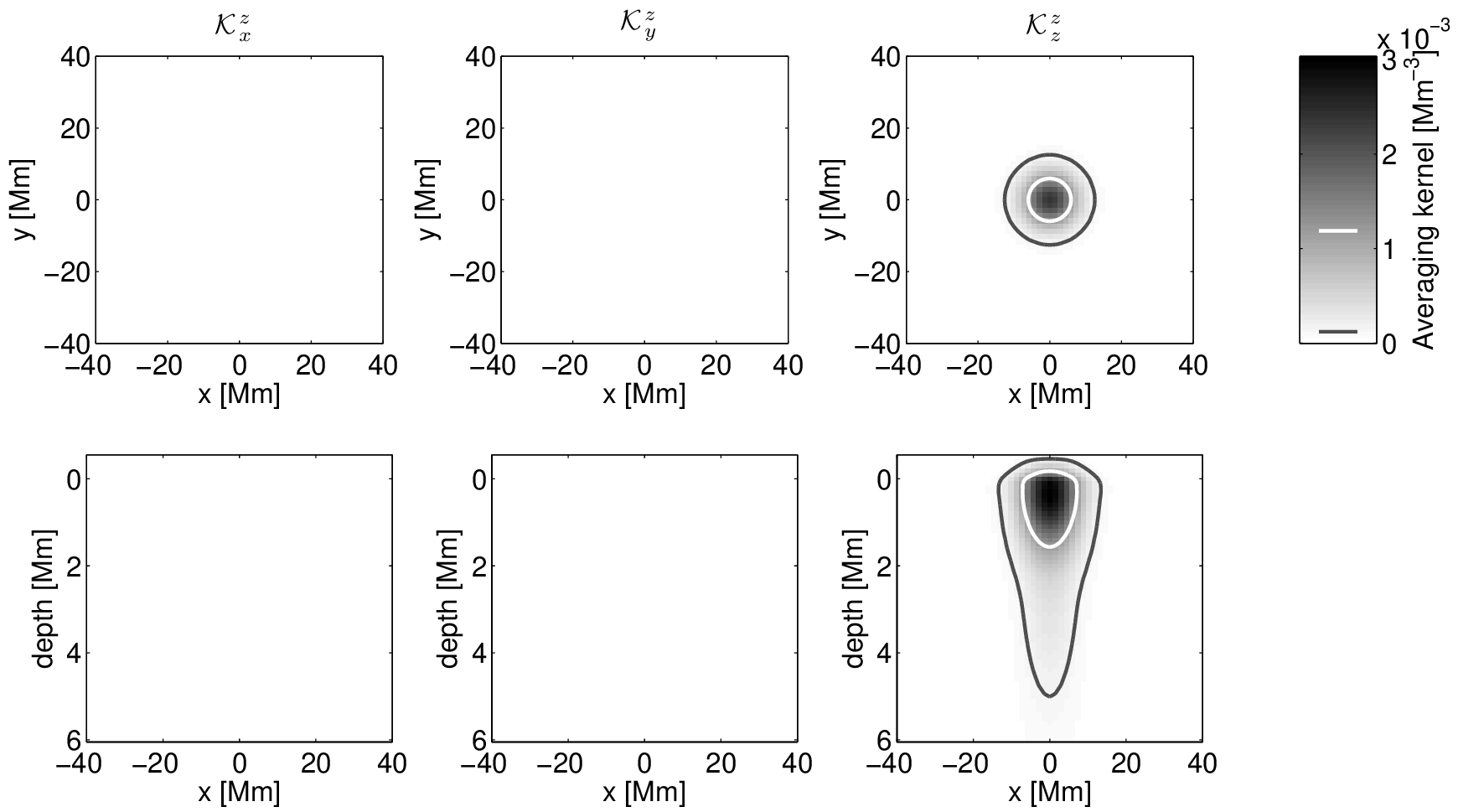}\\
\end{tabular}
\caption{Cuts through all components of averaging kernels from near-subsurface inversions for horizontal ($v_x$) and vertical ($v_z$) flows in average supergranule. 
Noise levels of these inversions are discussed in the main text. Over-plotted contours, which are also marked on the colour bar for reference, denote
the following: half-maximum of the kernel (white) and $5$\% of the maximum value of the kernel (grey). Components of the averaging kernels in the direction of the inversion ($\cK^x_x$ and $\cK^z_z$) are well localised without horizontal side-lobes. It is also evident that the contribution of the cross-talk is negligible as the cross-talk averaging kernels ($\cK^x_y$ and $\cK^x_z$ and $\cK^z_x$ and $\cK^z_y$) are practically zeroes. }
\label{fig:akerns}
\end{figure}

The averaging kernels of the discussed inversion are displayed in Fig.~\ref{fig:akerns}. It is clear that the flow estimates are averaged mostly over depths of 0--2~Mm with some contribution from deeper layers, especially in the case of the vertical flow inversion. The centre of gravity of the averaging kernels is located at a depth of 1.2~Mm.  

\section{An average supergranule}

The ensemble averaging approach was implemented to suppress the random-noise level and to form the set of travel-time maps representing an `average 
supergranule'. The supergranules were stacked about locations of their centres. The locations of centres of individual supergranules were determined using an algorithm similar to that used by
\cite{2010ApJ...725L..47D} and \cite{2012SoPh..tmp..136D}. The supergranules were searched for in 
the centre-to-annulus difference travel-time maps measured from $f$-mode filtered datacubes for the 
distance range of 8--12 pixels (0.96--1.44$^\circ$). After smoothing with a 
Gaussian window with Full-Width-at-Half-Maximum of 3~Mm to suppress random fluctuations, such maps were sensitive to the horizontal 
divergence of the flow, which is assumed to occur inside supergranular cells. 

The pixels with highly negative travel time (which corresponds to large divergence signal) were identified in each map. This identification resulted a list of potential cell centres. In the next step, the list was cycled and the distance to all other potential cell centres was calculated. Should the distance between any pair of points be less than 23~Mm, the point with weaker signal was removed from the list. In such way duplicities were removed. 

In all 64 datacubes, the procedure identified 5582 supergranule centres.

\subsection{Near-subsurface flows in the average supergranule}
For the inverse modelling, maps with independent realisations of travel times were averaged about the location of 
the supergranular centres. This resulted in a single set of travel-time maps, representing travel-time shifts 
caused by average anomalies under the average supergranule. The average travel-time maps were inverted for all flow components. The averaging over 5582 supergranules 
reduced the noise level by a factor of $\sqrt{5582}=76$ and led to the level of random noise in velocity estimates less 
than 2~\mps{}. 

\begin{figure}[!th]
\centering
\includegraphics[width=0.8\textwidth]{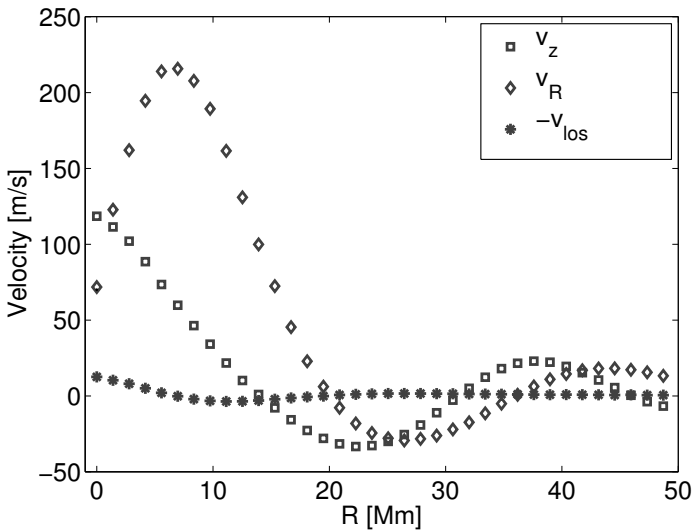}
\caption{Azimuthal average of the radial (with respect to the cell centre, diamonds) and vertical ($v_z$, squares) near-subsurface velocity in the average supergranule. The line-of-sight velocity ($v_{\rm los}$ with negative sign; full circles) obtained from the stacked Dopplergrams of supergranules is over-plotted for reference. It is practically identical to that presented in \cite{2010ApJ...725L..47D}. Error bars for all plots are 2~\mps{} or less. }
\label{fig:radialflows}
\end{figure}

The magnitude of the vertical and horizontal flow, as a function of the distance $R$ from the centre 
of the average supergranule, is plotted in Fig.~\ref{fig:radialflows}. The vertical velocity peaks in the cell 
centre with a value of 117~\mps. The root-mean-square vertical velocity over the area of the average supergranule is 21~\mps, 
well within the measurement by \cite{2002SoPh..205...25H}. The 
second peak in the vertical velocity, at a distance of 38~Mm from the cell centre, indicates the 
average location of the neighbouring supergranules in my sample and agrees well with the scale-length determined from the velocity spectrum by other authors. The amplitude of the horizontal velocity components (in Fig.~\ref{fig:radialflows} only the radial component is plotted) is in agreement with literature. 

The known location of the supergranules allows one to also average the measured Dopplergrams and obtain the average line-of-sight velocity, which corresponds to the vertical 
velocity at the surface. My results are similar to those of 
\cite{2010ApJ...725L..47D}, i.e. 12$\pm$1~\mps{} in the cell centre, and are also plotted in Fig.~\ref{fig:radialflows}.

\subsection{Is this model of an average supergranule reasonable?}
The large-magnitude vertical flow resulting from the inversion is somewhat surprising, as this has not been seen before except for study of \cite{2012SoPh..tmp..136D}. Thus, one can ask the question, whether such solution to the problem is physically feasible. The check whether the mass conservation holds is one of the simplest. By splitting the continuity equation
\begin{equation}
\nabla \cdot \rho\bvec{v} = 0,
\end{equation}
where $\rho$ stands for density and $\bvec{v}$ for the full velocity vector, into the horizontal and vertical parts, one obtains
\begin{equation}
\rho(z) \nabla_h \cdot \bvec{v}_h + \frac{\partial \rho(z) v_z}{\partial z}=0.
\end{equation} 
To check whether the mass conservation holds, one can use the ratio of the two terms, which should be close to unity, and both terms should have opposite signs. Here density $\rho$ is assumed to vary only with depth $z$, horizontal variations are neglected. Operator $\nabla_h$ is the horizontal part of the divergence operator applied to the horizontal velocity vector $\bvec{v}_h=(v_x,v_y)$ and can be evaluated accurately in the wave-vector space. 
\begin{equation}
d_h(x,y)=\nabla_h \cdot \bvec{v}_h(x,y)= {\cal F}^{-1} \left[ \bvec{k} \cdot \tilde{\bvec{v}}_h(k_x,k_y)\right],
\end{equation}
where ${\cal F}^{-1}$ represents the inverse Fourier transform, $\bvec{k}=(k_x,k_y)$ is a horizontal wave-vector and $\tilde{\bvec{v}}_h$ a Fourier image of the horizontal velocity vector. 

The vertical derivative term can be roughly estimated from the values measured at the target depth and at the surface. The extent to which the mass conservation holds is evaluated from the equation
\begin{equation}
\mu=\frac{\left[\rho(z_0)v_z(z_0)-\rho(z_t)v_z(z_t)\right]/(z_t-z_0)} {\rho(z_t) d_h},
\label{eq:massconv}
\end{equation}
where $z_t=1.2$~Mm is a target depth and $z_0=0$~Mm at the surface. In the centre of average supergranule, i.e. at $(x,y)=(0,0)$, for the vertical component of the flow one uses value of  $v_z(0,0;z_0)=10$~\mps{}  \citep{2010ApJ...725L..47D}, which results from direct line-of-sight velocity measurements and thus is representative for vertical velocity at the surface. Lets take  $v_z(0,0;z_t)=117$~\mps{} at the target depth and $d_h(0,0)=83.9$~m/s/Mm. Densities are given by Model~S \citep{1996Sci...272.1286C}. The ratio $\mu$ should be close to unity if the mass conservation roughly holds. Indeed, in the case of here presented model, $\mu=1.16$. The vertical derivative of the density momentum is only 16\% larger than the ``compensating'' horizontal divergence of the density momentum. The best agreement is achieved when one assumes the target depth of 1.4~Mm, which, given the rather extended averaging kernel, is still a representative value. 

\begin{figure}[!t]
\centering
\includegraphics[width=0.75\textwidth]{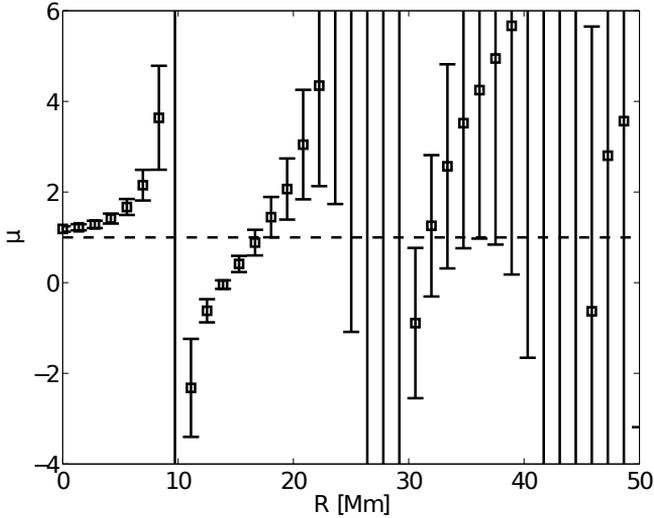}
\caption{An azimuthal average of the mass-conservation quantity $\mu$ defined by Eq.~(\ref{eq:massconv}) with 1-$\sigma$ error-bars as a function of the distance from the average cell centre $R$. In the ideal case, $\mu=1$ everywhere (dashed line).  }
\label{fig:massconv}
\end{figure}

The situation for the off-centre points is more difficult for two reasons. 
\begin{enumerate}
\item The averaging kernels for horizontal and vertical components are not identical, thus the inversion for independent components cannot be compared in a straightforward way.  
\item The representative of the surface vertical flow is taken directly from the stacked Dopplergrams. The difficulty with the latter can directly be seen in Fig.~\ref{fig:radialflows} -- $-v_{\rm los}$ turns negative much closer to the centre of the cell than $v_z(z_t)$. 
\end{enumerate}
\noindent The plot in Fig.~\ref{fig:massconv} shows that $|\mu| < 10$ within the full cell with a change of sign, which is correlated to the change of sign of the horizontal divergence of the density momentum. Given the roughness of the estimate and taking measurement error-bars into account, the mass conservation holds well.

\section{Concluding remarks}

The presented results can still be affected by various biases discussed below. 

\begin{itemize}
\item{\bf Detection of supergranules} The searching algorithm used in this study prefers the supergranules with highest divergence signals. It 
detected 5582 cells. Assuming that supergranules cover the whole Sun in a hexagonal pattern with a flat-to-flat distance of 30~Mm, around 50\,000 cells are expected. Thus, to obtain a better representative of average supergranule a different segmentation algorithm \citep[such as by][]{2008SoPh..251..417H} must be used, which detects not only the centres, but the boundaries as well. It also is not clear whether the large divergence signal belongs to supergranulation or to some smaller scale (and thus larger amplitude) convection mode. The algorithm based on pattern detection would overcome this issue too.  
\item{\bf Stacking of supergranules} The detected supergranules are stacked 
to the point of the maximal divergence, which does not necessary need 
to be in their geometrical centre. To overcome this issue, the detection code was modified to fit paraboloids in the vicinity 
of points with largest negative difference $f$-mode travel times and to choose vertices of these paraboloids as 
locations of the cell centres. The results remained largely unchanged. 
\item{\bf Travel-time measurement problem} Due to the complexity of the procedure, it is possible to determine 
travel times incorrectly, e.g., when one makes an error in the implementation of one of the steps necessary 
in the travel-time measurement. However, \cite{2012SoPh..tmp..136D} used a completely different travel-time code and a 
very different spatio-temporal filtering (phase-speed filters instead of 
ridge filters used here). Despite these significant differences in the implementation, they arrived basically at the same results.
\item{\bf Problem in the inversion code} The inverse problem for time--distance helioseismology is well 
described \citep{fastOLA}, but an error can be made in the implementation. It has to be noted that the 
same inversion code was properly and successfully validated using synthetic data \citep{Svanda2011}. 
\item{\bf Problem with sensitivity kernels} It is possible that the 
sensitivity kernels used in this study resulting from the forward 
modelling are not good enough, which would in turn affect the results of the inverse modelling. 
Again, \cite{2012SoPh..tmp..136D} did not perform the inverse modelling and thus did not use any 
sensitivity kernels in interpreting their results. And again, despite the crucial deviations in the approach, the results of their
study are qualitatively consistent with mine. 
\end{itemize}

\noindent There is a lot to be done to confirm the model of supergranule with large-amplitude flows. \cite{2012SoPh..tmp..136D} predicted also large-amplitude horizontal flows (some 700~\mps{} at the depth of 1.6~Mm). Only a very detailed tomography with a fine depth resolution may answer related questions.

\section*{Acknowledgements} 
I acknowledge the support of the Czech Science Foundation\break (grant P209/12/P568) and of the Grant Agency of Academy of 
Sciences of the Czech Republic (grant IAA30030808). This work utilised the resources and helioseismic products 
dispatched within the German Science Center at MPS in Katlenburg-Lindau, Germany, which is supported by German 
Aerospace Center (DLR). The data were kindly provided by the HMI consortium. The HMI project is supported by NASA contract NAS5-02139. 
Tato pr\'ace vznikla s podporou na dlouhodob\'y koncep\v{c}n\'\i{} rozvoj v\'yzkumn\'e organizace (RVO:67985815).

\bibliographystyle{ceab}
\bibliography{inversions}


\end{document}